# Shallow lunar fractures dominate the long-lasting seismic coda

Lei Zhang[1], Qian Huang[2], Minghao Huang[1], Jinhai Zhang[1,*], Dijun Guo[3]

[1] Institute of Geology and Geophysics, Chinese Academy of Sciences, Beijing, China.

[2] Hubei Subsurface Multi-scale Imaging Key Laboratory, School of Geophysics and Geomatics, China University of Geosciences, Wuhan, China.

[3] State Key Laboratory of Solar Activity and Space Weather, National Space Science Center, Chinese Academy of Sciences, Beijing 100190, China.

*Corresponding to Jinhai Zhang (zjh@mail.iggcas.ac.cn)

**Lunar seismic records are characterized by long-lasting spindle-shaped coda, which has long been considered to be caused by strong velocity contrasts in a continuous-medium lunar crust; however, simulated waveforms could not well match the observation even with unrealistically strong velocity contrasts, demonstrating its mechanism still unsolved. Here we propose a novel model of coda wave generation with random fractures in the lunar crust, where vacuum fractures prevent seismic wave propagation as insulating boundaries, causing strong persistent reverberation in terms of total reflection among fractures. Our model can easily reproduce the typical characteristics of observed Apollo data even using a limited number of simplified fractures in homogeneous matrix, with envelope correlations over 0.9, much higher than previous studies. Numerical simulations suggest that fractures in the uppermost ~1 km of the crust may dominate lunar seismic scattering. This model enables us to investigate the shallow lunar fracture system using seismic data and to suppress seismic coda for detecting critical phases reflected from deep interiors.**

## INTRODUCTION

Seismic observations provide a primary way of probing lunar interiors[1,2], but the identification of seismic phases is greatly challenging due to three main obstacles: a limited number of available waveforms, a small magnitude of moonquakes, and an unusually long-lasting seismic coda[3,4]. The lunar seismic coda can last for up to hours, about ten times longer than for an earthquake with a similar magnitude[5]. This greatly hinders the identification of critical deep lunar reflections, limiting constraints on the spherical structures[6,7]. It is therefore critical to investigate the physical origin of lunar seismic coda, which is also important to improve seismic phase identification for successfully probing deep lunar interiors.

Numerous studies attributed seismic coda to the seismic scattering effect caused by spatial variations of crustal heterogeneous, based on continuous-medium models[8-13]; however, their simulated waveform envelope could not well match the Apollo observations, even with unrealistically strong velocity contrasts. This indicates that the continuous-medium models could not well explain the coda wave generation, even though strong heterogeneity may contribute to somewhat.

Many studies show that the lunar crust contains pervasive fractures and porosity due to prolonged impact bombardment[14-16], which may exhibit in different scales. Besides, cooling-driven thermal contraction provides persistent force to generate faults and fractures, while tidal stress and volcanic deformation can further activate crustal fractures[17-20]. However, fluid-mediated fracture healing, although common on Earth, could not happen due to the absence of water on the Moon[12,21-24]. As a result, the fractures in the lunar crust would persist till today (Fig. 1), significantly distorting seismic wave propagations. Although the influence of terrestrial fractures has been noticed[25], the effect of fractures on lunar seismic wave propagation has never been considered, to the best of our knowledge.

Here, we propose a discontinuous model (Methods) to account for the influence of vacuum fractures within the lunar crust on seismic coda waves. First, we develop a full-waveform simulation framework to explicitly incorporate shallow fracture networks into the lunar crust, based on the spectral-element method[26-28]. Then, we estimate the geometric feature of fractures and their effective depth by comparing the simulated waveforms with Apollo observation (Methods). Finally, we discuss the possible formation mechanisms of lunar fractures and their implications for probing the lunar interior.

## RESULTS

### Modelling lunar fracture scattering with Apollo impact records

To test whether shallow fractures can explain the unusually long lunar seismic coda, we construct a 2D spectral-element model (Extended Data Fig. 1) of the lunar crust (Fig. 1). The model contains vacuum fractures in a homogeneous background medium, with the fracture distribution guided by regional gravity and geomorphic observations (Fig. 1c–g)[14,22,29,30]. We then test the model against 13 Apollo artificial-impact records spanning epicentral distances of 2.2°–28.0° (Fig. 1a,b, Extended Data Table 1)[31]. These records are well suited for testing the scattering mechanism because their source locations and origin times are well constrained with high signal-to-noise ratios[13]. We compare the observed and simulated waveform envelopes only within the most reliable frequency band (0.3–0.8 Hz) of the observed data[1,4,13, 32].

**Shallow fracture scattering explains the long Apollo seismic coda**

To understand how fractures generate a long-lasting seismic coda, we compare wave propagation in the continuous-medium model (no fracture) and the fractured model (with a single fracture or multiple fractures) (Fig. 2a–i). In the continuous-medium model, wavefronts always propagate coherently (Fig. 2a–c). Before the wave reaches the fracture, the wavefield in single-fracture model remains almost the same as that in the non-fracture model (Fig. 2d,e); in contrast, we can observe clear reflections after the wave reaches the fracture, with most parts of the wavefront undisturbed (Fig. 2f), where the fracture acts as a total reflection boundary, with no any transmissive wave across the fracture. For the multiple-fracture model (Fig. 2g-i), the wavefront would be completely broken up and trapped due to too many times of total reflections among fractures, eventually acting as strong scattering effect (Fig. 2h,i). These results show that repeated total reflection within a shallow fracture network spreads seismic energy over a much longer time, producing an extended lunar coda.

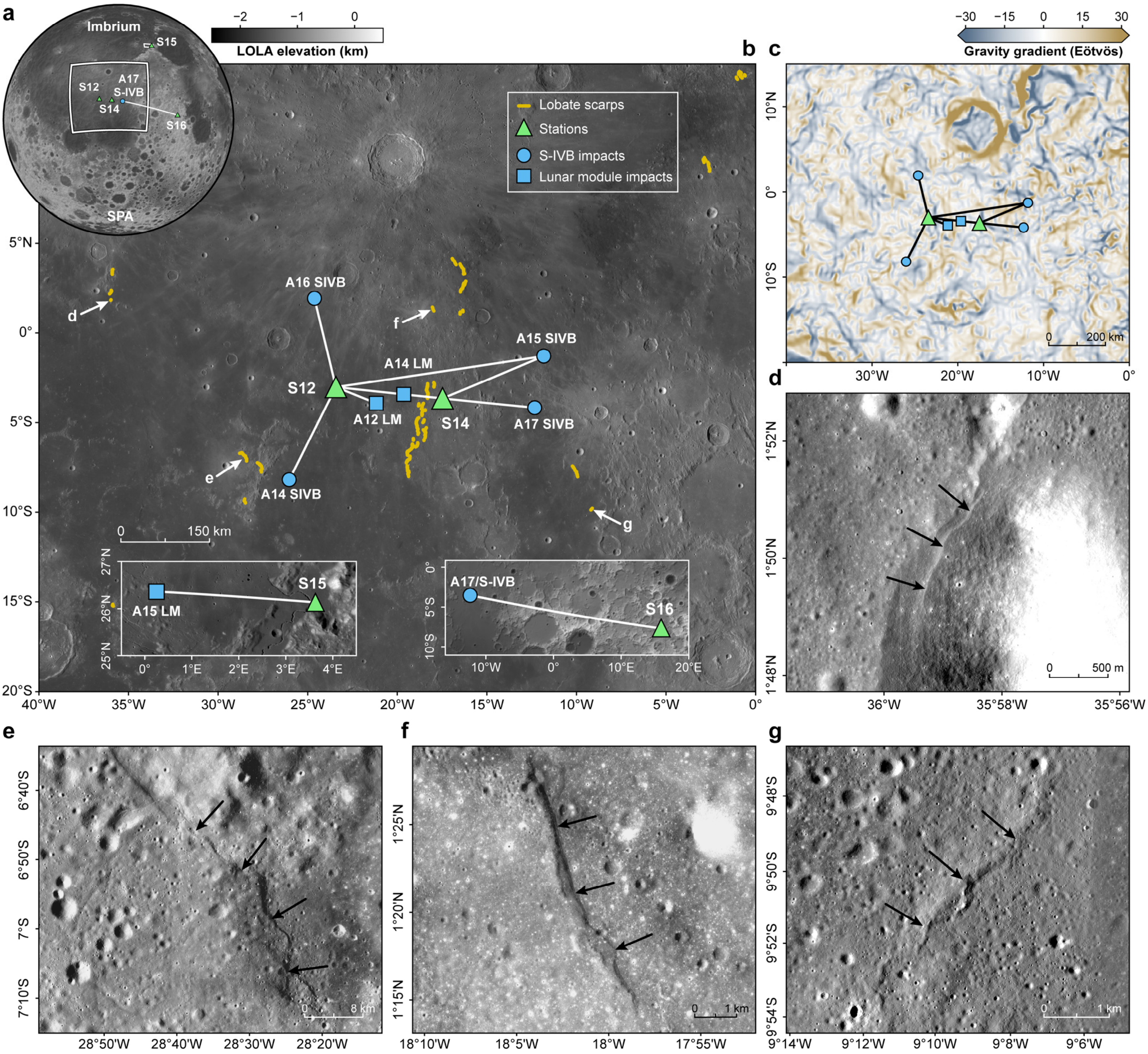

**Fig. 1 | Geological and gravity context of the Apollo study region. a**, Global context from LOLA shaded relief, where the white box outlines the study region. **b**, The Apollo landing region from the LROC WAC global morphology mosaic (100 m/pixel), showing the seismic stations S12, S14, S15 and S16 as green triangles, the artificial impacts as squares and circles, the source–receiver paths as white lines, and mapped lobate scarps in yellow[30]; the insets show the A15 and A17 paths; white arrows mark panels d–g. **c**, GRAIL Bouguer gravity-gradient map of the same region with the seismic paths overlain, providing regional context for crustal density heterogeneity. **d**–**g**, LROC NAC close-ups (~3 m/pixel) of representative lobate scarps at the locations boxed in b.

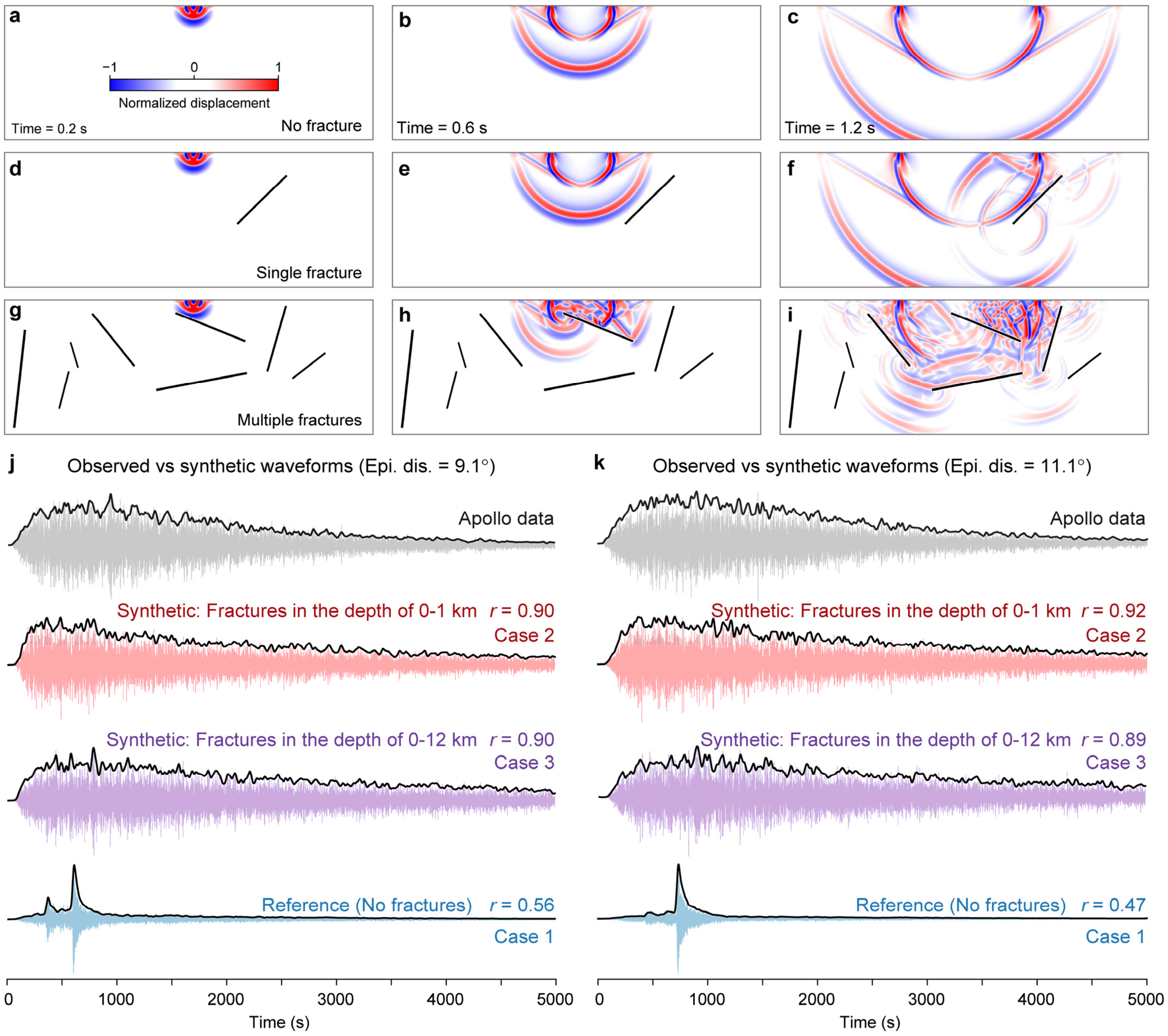

**Fig. 2 | Shallow fracture scattering generates the Apollo coda. a–i**, Wavefield snapshots at $t = 0.2$, 0.6 and 1.2 s for the continuous model (a-c), and the fractured model with a single fracture (d-f) or many fractures (g-i); black lines indicate fractures. These local-scale simulations use a 10-Hz source, and the length of single fracture in d–f is 500 m. **j,k**, Observed Apollo waveforms (gray) at epicentral distances of 9.1° (j) and 11.1° (k), compared with synthetics for fractures in the top 0–1 km (red), fractures through 0–12 km (purple) and the continuous model (blue); Black curves show the corresponding smoothed amplitude envelopes. $r$ indicates Pearson envelope correlation over 0–4500 s.

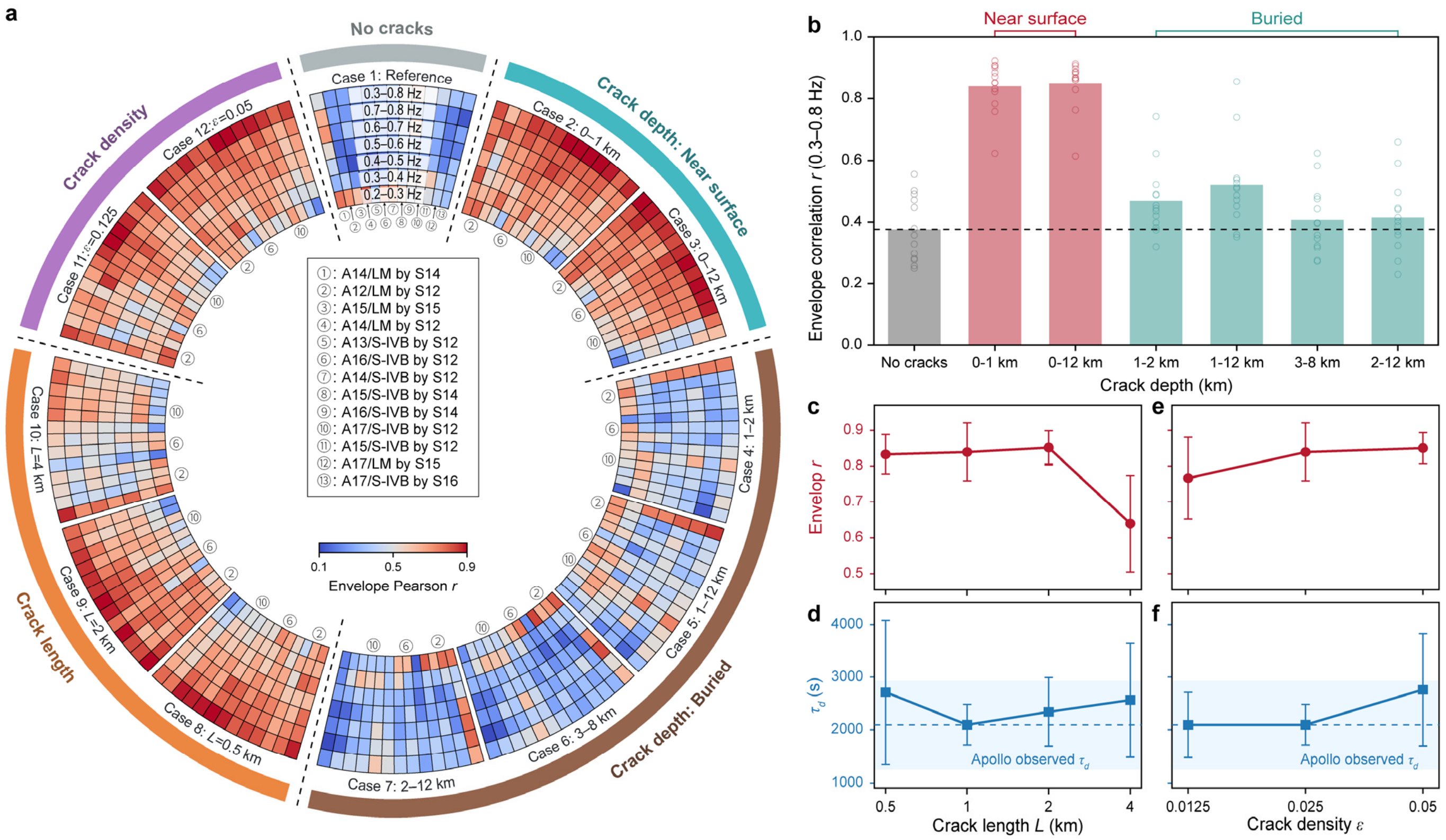

**Fig. 3 | Sensitivity of coda fits to fracture depth, length and density across 12 models.** **a**, Twelve crustal scenarios (Cases 1–12) in five groups: no fractures; surface-reaching fractures (0–1 and 0–12 km); buried fractures (1–2, 1–12, 3–8 and 2–12 km); fracture length ($L$ = 0.5–4 km); and fracture density ($\varepsilon$ = 0.0125–0.05). Each wedge shows the envelope correlation $r$ across seven frequency bands (0.2–0.8 Hz) and 13 impact–station pairs at epicentral distances of 2.2–28.0°; circled numbers identify the records listed in the central legend. **b**, Envelope correlation $r$ (0.30–0.80 Hz) versus fracture depth range, contrasting surface-reaching (red) with buried (teal) fractures and the continuous model (grey); open circles, individual records; dashed line, the continuous-model mean. **c**,**d**, Envelope correlation $r$ (red, c) and coda decay time $\tau_d$ (blue, d) versus fracture length $L$; dashed line and shading, observed value and range. **e**,**f**, As in c,d, but versus fracture density $\varepsilon$. Error bars in c–f indicate ±s.d. across records.

We next test whether this fracture-scattering mechanism can reproduce the observed shape and duration of the Apollo seismic coda. For the two representative events (Fig. 2h,i), when the fractures concentrated in the uppermost kilometer, the simulated waveforms can simultaneously match the main features of the observed spindle-shaped envelopes: a gradual rise in amplitude, a broad peak, and a slow decay. Over the time window of 0–4500 s, the correlation coefficient between the observed and simulated envelopes are 0.90 or even 0.92 (Fig. 2j,k), much higher than that of 0.56 and 0.47 for the continuous-medium models. These results suggest that shallow fracture networks would trap seismic energy by repeated reflections, promoting the gradual build-up and the slow decay of seismic coda.

**The uppermost kilometer fractures dominate lunar coda scattering**

We further evaluate the contribution from the fracture depth, length, and density that may affect the seismic coda (Fig. 3a; Methods). The envelope fits are particularly sensitive to the presence of fractures in the uppermost kilometer. Models with fractures confined to 0–1 km (Case 2) or extending to 12 km depth (Case 3) can best match the observations (Fig. 3b; Extended Data Fig. 2), with correlation coefficient up to 0.92; in contrast, the correlation coefficient is only 0.4–0.5 when removing fractures from the uppermost kilometer, demonstrating that the uppermost kilometer fractures dominate lunar coda scattering. When the fracture length is 0.5–2 km, the simulated results can obtain high correlation coefficients of ~0.85, while 4-km fractures can only reach ~0.64 (Fig. 3c), demonstrating that the fracture length may be ~2 km or shorter. When the fracture density is relatively high, we can obtain acceptable fits on both envelope correlation and coda decay time ($\tau_d$), demonstrating that the uppermost 1-kilometer is highly fractured, consistent with theoretical analyses[25] (Fig. 3e,f; Extended Data Fig. 3).

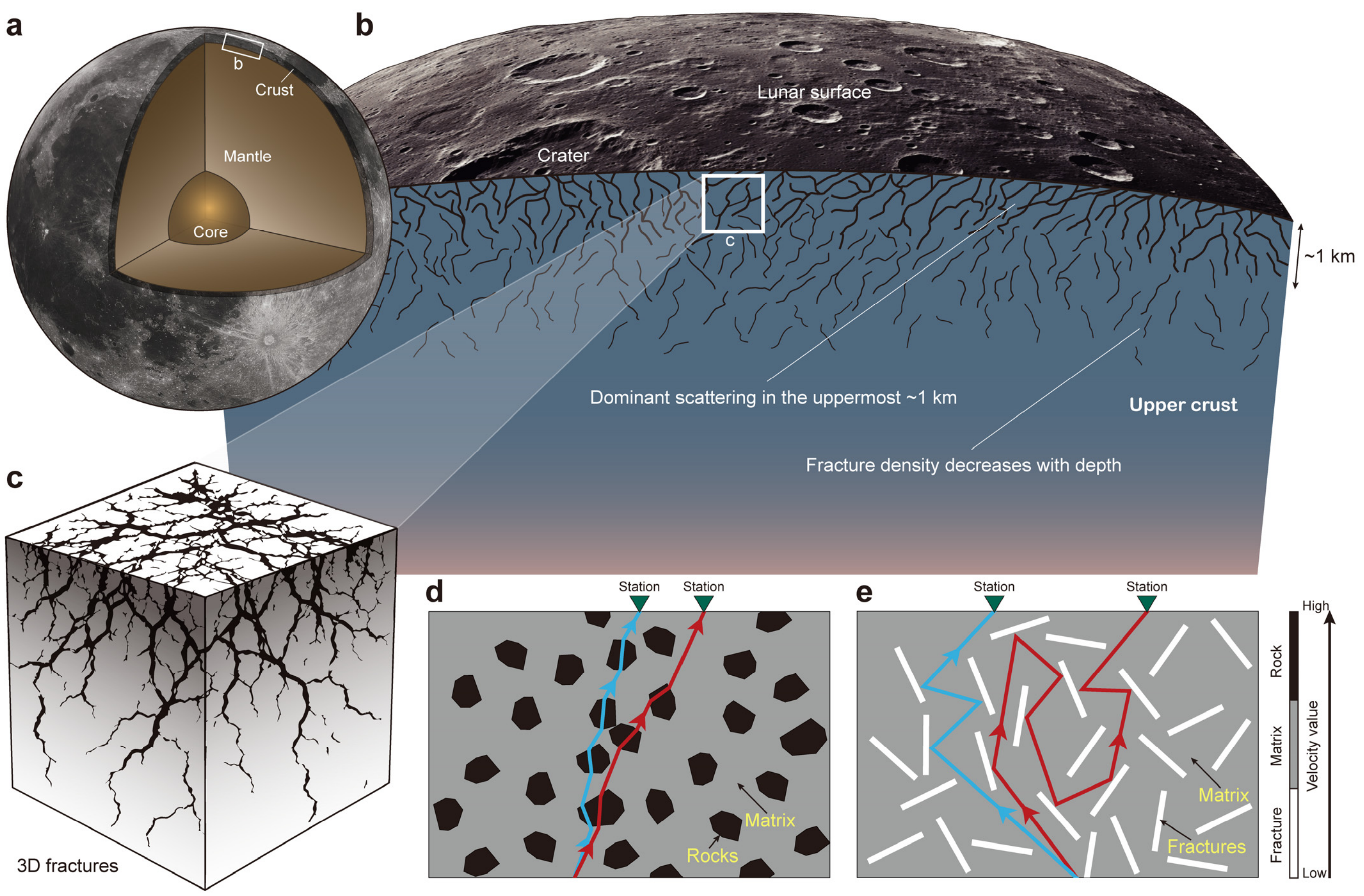

**Fig. 4 | Conceptual model of shallow lunar fractures and their effects on seismic-wave propagation**. **a**, Schematic lunar interior showing the location of the shallow fractured crust. **b**, Cross-section of the lunar crust showing a dense fracture network near the surface, with fracture abundance decreasing with depth. **c**, Enlarged three-dimensional view of the shallow fracture network. **d**,**e**, Comparison of seismic-wave propagation through distributed rock heterogeneity (d) and vacuum fractures (e). The red and blue lines represent wave paths toward a seismic station. Unlike distributed rock heterogeneity, vacuum fractures produce strong reflections and multiple scattering, increasing the complexity of wave propagation and contributing to the prolonged lunar seismic coda.

## Discussions

### 1. On the fracture model

Previous studies have generally modeled lunar scattering using continuous heterogeneous or stochastic velocity structures[12,13,33-37]. In contrast, our model here uses a homogeneous background medium for simplicity but with strong scattering generated by vacuum fractures (Fig. 4d,e; Methods). The close agreement achieved with this simple setup highlights the importance of near-surface vacuum fractures in controlling lunar seismic scattering. More realistic models can build on this framework by incorporating fractures into heterogeneous crustal structures and jointly considering other sources of scattering, such as distributed velocity heterogeneity, porosity variations, and regolith layering. Extending the model to three-dimensional fracture networks, although much more complicated in numerical description and implementation[26], would better represent fracture orientations, intersections, and wave interference in 3D scenario, thereby better matching the observations.

The scattering behavior identified here provides a forward-modeling basis for quantitative characterization of lunar fracture system. By fitting waveform envelopes and coda-decay trend, this approach can be extended to constraining the depth range, characteristic length, density, and spatial distribution of shallow fractures. Synthetic waveforms generated from constrained fractured models could also guide the suppression or separation of shallow-scattering energy from lunar seismic records, enhancing the detectability of weak phases that are reflected or converted at deep interfaces.

## 2. Implication on multidisciplinary constraints

Observations at different spatial scales provide complementary evidence for crustal damage and depth-dependent porosity in the shallow lunar crust. Chang'e-3 and Chang'e-4 lunar penetrating radar measurements reveal layered and impact-modified structures extending to several hundreds of meters depth[38,39]. GRAIL observations indicate a porous and heterogeneous crust with porosity generally decreasing with depth[14,22,40,41]. Analysis of Chang'e-6 returned samples, combined with geophysical and compaction modeling, further supports a strongly depth-dependent porosity structure[14,42,43]. Although these results indicate crustal damage and variations in porosity with depth, they could not constrain the depth range of fractures. Our fracture-based seismic modeling provides this missing constraint by identifying the uppermost ~1 km as the dominant scattering region, which is deeper than the maximum measuring depth (typically <500 m) of ground penetrating radar. Our seismic results can bridge the radar detections, gravity observation, and sample-derived models of crustal density and porosity. Integrating these constraints into joint geophysical inversions could improve the characterization of lunar crustal structure.

## 3. Implications for crustal evolution and gravitational-wave detection

The long lunar seismic coda may contain information about the accumulated damage and evolution of the shallow lunar crust. Over billions of years, impact bombardment, thermal contraction, and crustal modification have shaped the lunar crustal fracture system, while subsequent burial and compaction may have reduced fracture abundance at depth. These fractures not only control seismic scattering but may also influence crustal heat transport by changing the effective thermal conductivity of rocks[44-46]. The depth distribution of fractures inferred from seismic observations could therefore help understanding how these geological processes have altered the physical and thermal properties of the lunar crust.

Future lunar seismic missions, including Chang'e-7 and Artemis-related seismic deployments near the lunar south pole, as well as the Farside Seismic Suite on the lunar farside[47-49], could independently test our fractured model and extend its application beyond the Apollo landing sites. Recently, several works start to discuss the possibility of detection gravitational waves on the Moon[27,28,50-53]; whereas, the potential influence of fractures has not been considered. Our model can help to test its effect thus improves the calibration of gravitational-wave measurements.

## Methods

### 1. Apollo seismic data preprocessing

#### 1.1 Original sources of Apollo seismic data

To compare simulated and observed seismic envelopes, we selected 13 artificial-impact event–station pairs recorded at the Apollo 12, 14, 15 and 16 stations (Fig. 1; Extended Data Table 1). These signals were generated by impacts of Saturn V (S-IVB) rocket stages and lunar module (LM) ascent stages, whose known source locations and impact times reduce uncertainty in propagation geometry and timing. Waveforms and station-response data for network XA (https://doi.org/10.7914/SN/XA_1969) were obtained using ObsPy[54,55]. We used the vertical mid-period component (MHZ) except for the Apollo 16 S-IVB impact recorded at the Apollo 14 station, for which only horizontal components were available; MH1 was selected for this event following ref. 13.

#### 1.2 Timing correction and despiking

To establish a unified time reference, we corrected sampling-clock errors following ref. 31 (ATT timing channel and PDART), calibrating the time axis to the original impact time, defined as $t$=0. Isolated single-sample gaps were repaired by linear interpolation, while longer gaps were retained as missing values and excluded from subsequent analyses.

The original Apollo records contain many spikes. To reduce their influence on waveform and envelope estimates, we used PDART to remove isolated spikes and replace affected samples by linear interpolation[31]. Residual anomalous intervals were identified by manual inspection and masked as NaN values, excluding them from subsequent filtering, envelope calculation and correlation analysis (Extended Data Fig. 4).

### 1.3 Instrument-response removal and filtering

To obtain displacement records comparable with the simulations, we applied instrument-response removal and frequency-domain prefiltering using ObsPy. Each continuous data segment was processed independently, with prefilter corner frequencies of [0.03, 0.05, 0.90, 1.10] Hz. Each segment was demeaned and tapered at both ends, with the taper duration on each side limited to the smaller of 5% of the segment length and 20 s.

To place observations and simulations on a unified time step, we resampled the displacement records to 4 Hz using polyphase FIR resampling with anti-alias filtering implemented in SciPy (https://scipy.org/). Resampling was applied independently within each continuous valid segment, preserving long gaps and manually masked intervals.

## 2. Model construction and parameter setting

### 2.1 Model geometry and mapping of Apollo paths

For each Apollo artificial-impact event, we placed the source at the same location in the two-dimensional lunar crustal model and positioned the receivers according to their source–station distances. Receiver positions represented the epicentral distances of the 13 selected

Apollo records, spanning 2.2°–28.0° (Extended Data Table 1). The model used a homogeneous background medium so that the effects of discrete vacuum fractures could be evaluated. The modeled section extended 1000 km along the surface and 80 km in depth.

### 2.2 Background structure and fracture distribution

The background structure was based on a modified VPREMOON model[32], incorporating core parameters from ref. 1. The model parameters are provided in https://github.com/zhanglei911/LunarFracture. Fractures were represented as rectangular voids within this background medium. Their depth intervals, lengths ($L$) and densities ($\varepsilon$) were varied among the fractured models (Extended Data Table 2). Case 1 contained no fractures and served as the reference model.

Following the two-dimensional convention of ref. 25, fracture density was defined as

$$\varepsilon = \nu a^2 = \frac{NL^2}{4A},$$

where $N$ is the number of fractures, $\nu = N/A$ is their number density and $a = L/2$ is the fracture half-length. The reference area $A$ is the full model surface arc length multiplied by the depth range. For the shallow reference configuration (Case 2), a minimum centre depth of 150 m and a 1-km lower boundary gave $A = 1000 \times 0.85 = 850$ km². The 85 fractures of full length $L = 1$ km therefore gave $\varepsilon = 0.025$. Their width was 250 m, giving a length-to-width ratio of 4.

Candidate fracture-centre depths were sampled from an exponential distribution truncated to the prescribed depth interval, consistent in form with exponential porosity profiles adopted in previous lunar crustal models[14,42]. Candidate orientations were sampled uniformly over 0–180° in model coordinates. Rejection sampling ensured a minimum edge-to-edge fracture spacing of 1 km.

### 2.3 Mesh generation and simulation settings

Following the modelling approach of refs. 26-28, we simulated wave propagation using the spectral-element code SPECFEM2D[56,57]. The unstructured interior mesh was generated in Gmsh and recombined into quadrilateral elements, with finer elements around fractures and within the shallow fractured layer and coarser elements at greater depth (Extended Data Fig. 1b–d). Perfectly matched layers (PMLs), 6 km thick, were placed along the lateral and bottom boundaries to absorb outgoing waves (Extended Data Fig. 1). Meshes for the other cases were generated using the same procedure, with local manual adjustments where needed to reduce element distortion. The time step was 0.0004 s. The source-time function was a Ricker wavelet with a dominant frequency of 0.5 Hz. Separate local-scale simulations in Fig. 2a–i use a 10-Hz source to illustrate wave–fracture interactions; the length of single fracture in Fig. 2d–f is 500 m.

### 2.4 Model cases and sensitivity analysis

The 12 model cases were organized into five groups (Extended Data Table 2): the continuous model (Case 1), surface-reaching fractures (Cases 2–3), buried fractures (Cases 4–7), variations in fracture length (Cases 8–10), and variations in fracture density (Cases 11–12). Case 2 served as the baseline fractured model, with fractures in the uppermost 1 km, $L = 1$ km, and $\varepsilon = 0.025$. Case 3 extended the fracture distribution to 12 km, while Cases 4–7 placed fractures at depths of 1–2, 1–12, 3–8, and 2–12 km, respectively, with $L = 1$ km and $\varepsilon = 0.025$. Cases 8–10 used $L = 0.5$, 2, and 4 km at $\varepsilon = 0.025$; their fracture depth intervals were 0–1, 0–2.5, and 0–4.5 km, respectively. The depth intervals in Cases 9–10 were extended to accommodate the longer fractures. Cases 11–12 retained the depth interval and fracture length of Case 2 and used $\varepsilon = 0.0125$ and 0.05. In addition to these 12 cases, the fracture-width parameter was varied by ±50% relative to Case 2 to evaluate its effect on correlations between observed and synthetic envelopes (Extended Data Fig. 5).

**3. The SPECFEM2D simulation result preprocessing**

Using the SPECFEM2D, we output horizontal (BXX) and vertical (BXZ) displacements in the Cartesian coordinates, respectively. Then we projected these components onto local radial and tangential directions defined by each station position and the center of the lunar cross section. The radial displacement was compared with Apollo MHZ records, except for the record of the Apollo 16 S-IVB impact at the Apollo 14 station, for which tangential displacement was compared with the MH1 record. Simulated records were resampled to 4 Hz using the same polyphase FIR resampling and anti-alias filtering as the observations.

**4. Envelope and correlation analysis of observed and simulated signals**

## 4.1 Hilbert transform

To compare the shapes of observed and simulated seismic envelopes, we calculated the envelope of each bandpass-filtered displacement record. We processed the continuous record segments separated by data gaps or masked anomalous intervals independently, removed the mean and linear trend, and used a fourth-order, zero-phase bandpass filter. The envelope was extracted using the Hilbert transform[58] as:

$$E(t) = |z(t)| = \sqrt{x_f^2(t) + \{H[x_f(t)]\}^2}$$

where $t$ is time in seconds, $x_f(t)$ is the displacement waveform filtered within the selected frequency band, and $\mathcal{H}[x_f(t)]$ is its Hilbert transform. $E(t)$ denotes the unsmoothed amplitude envelope and has the same units as displacement. Amplitude envelopes characterize the temporal evolution of scattered coda while suppressing the rapid oscillations of the original waveforms[59].

## 4.2 Envelope Gaussian smoothing

To suppress rapid envelope fluctuations and compare shape, we smoothed the envelopes by convolution with a Gaussian kernel using SciPy. For a frequency band bounded by $f_{\min}$ and $f_{\max}$ in hertz, the centre frequency is $f_c = \frac{f_{\min}+f_{\max}}{2}$. The kernel full width at half maximum (FWHM) and standard deviation were set to

$$T_{\mathrm{FWHM}} = \max(25\ \mathrm{s}, \frac{10}{f_c}), \sigma = \frac{T_{\mathrm{FWHM}}}{2\sqrt{2\ln 2}}.$$

Both $T_{\mathrm{FWHM}}$ and $\sigma$ are expressed in seconds. This sets the FWHM to ten centre-frequency periods with a minimum of 25 s. The kernel was truncated at $\pm 4\sigma$, normalized to unit sum and applied by FFT convolution independently within continuous valid segments, with mirror extension at both ends to reduce boundary artifacts. Sensitivity to the smoothing width was assessed by comparing envelope shapes and observed–synthetic correlations under different settings (Extended Data Fig. 6).

### 4.3 Observed–synthetic envelope correlation

Envelope correlation provides a measure of similarity in temporal envelope shape[60]. To evaluate how well the fracture models reproduce the Apollo observations, we calculated Pearson correlation coefficients between the Gaussian-smoothed observed and simulated envelopes. Both records were resampled at 4 Hz, corresponding to a sampling interval of 0.25 s, and aligned at their respective original impact and numerical source times. Correlations were evaluated over 0–4500 s at zero lag.

The envelope correlation coefficient was calculated as

$$\boldsymbol{r} = \frac{\sum_{i=1}^{N}(O_i-\bar{O})(S_i-\bar{S})}{\sqrt{[\sum_{i=1}^{N}(O_i-\bar{O})^2][\sum_{i=1}^{N}(S_i-\bar{S})^2]}}.$$

Here, $O_i$ and $S_i$ are paired Gaussian-smoothed observed and simulated envelope values, $N$ is the number of valid pairs, and the overbars denote their means. $\bar{O}$ and $\bar{S}$ are the arithmetic means calculated over those same pairs. The coefficient ranges from −1 to 1, with higher values indicating closer agreement in temporal evolution.

To examine the frequency dependence of model performance, we applied identical processing to observations and simulations in six frequency bands: 0.20–0.30, 0.30–0.40, 0.40–0.50, 0.50–0.60, 0.60–0.70 and 0.70–0.80 Hz and a broader 0.30–0.80 Hz band. Cases 2 and 3 were selected as representative models and the synthetic envelopes were compared with observations for all 13 impact–station pairs (Extended Data Fig. 2).

**5. Estimation of coda decay time**

To compare coda decay between simulations and observations, we extended the envelope-based regression method of ref. 13 from the 0.40–0.50 Hz band to a broader 0.40–0.80 Hz band, using smoothed displacement-amplitude envelopes.

Envelopes were smoothed using a centered moving average with a half-window of 150 periods at the center frequency (0.60 Hz), equivalent to 250 s and a total window of approximately 500 s. The envelope peak time, $T_{\max}$, was identified within 0–4500 s. We fitted an exponential decay over a 1000 s window beginning at $t_s = T_{\max} + 200\,\mathrm{s}$ by ordinary least-squares regression in logarithmic-amplitude space:

$$\tilde{E}(t) = Ae^{\frac{-t}{\tau_d}}$$

where $\tilde{E}(t)$ is the smoothed amplitude envelope, $A$ is the fitted amplitude scale factor and $\tau_d$ is the time required for the fitted amplitude envelope to decrease to $1/e$ of its initial value. Parameters were estimated by ordinary least-squares regression of logarithmic envelope amplitude against time (Extended Data Fig. 3).

Estimates of $\tau_d$ are relatively sensitive to parameter choices[13]. We therefore held the frequency band, smoothing half-window, post-peak delay and fitting duration fixed across observations and simulations.

## Data availability

The LROC WAC global morphology mosaic and NAC images are available from the USGS Astrogeology Science Center (https://astrogeology.usgs.gov/search/map/moon_lro_lroc_wac_global_morphology_mosaic_100m) and the LROC archive (https://www.lroc.asu.edu), respectively. GRAIL gravity data are available from the NASA Planetary Data System (https://pds-geosciences.wustl.edu/missions/grail/). Apollo seismic waveforms and instrument-response metadata are available from the IRIS Data Management Center under network XA (1969–1977; https://doi.org/10.7914/SN/XA_1969). Data for a representative numerical simulation are archived at Zenodo (https://zenodo.org/records/23503054).

## Code availability

The codes used in this study are available to interested researchers upon request. The code and associated data for a representative numerical simulation of seismic-wave scattering by fractures in the lunar crust are archived at Zenodo (https://zenodo.org/records/23503054) and are also available on GitHub at https://github.com/zhanglei911/LunarFracture.

## Acknowledgements

We thank the LROC, GRAIL and Apollo mission teams for making their data publicly available, and the USGS Astrogeology Science Center, NASA Planetary Data System and IRIS Data Management Center for providing access to the datasets used in this study. We also thank the developers of Gmsh, ObsPy and PDART for making their tools for mesh generation and seismic data processing publicly available.

## Author contributions

Conceptualization: J.Z. and L.Z. Methodology: L.Z., M.H., and J.Z. Interpretation: L.Z., J.Z., Q.H., and D.G. Investigation: L.Z. and J.Z. Visualization: L.Z., M.H., J.Z., Q.H., and D.G. Supervision: J.Z. Writing—original draft: L.Z. and J.Z. Writing—review & editing: All authors.

## Funding

This work was supported by the National Natural Science Foundation of China Grant (Nos. 42622802, 42204178 and 42472304), and the National Key R&D Program of China (2021YFA0715100).

## Competing interests

The authors declare no competing interests.

**Extended Data figure/table legends**

**Extended Data Table 1 | Apollo signals used from Saturn IVB booster and Lunar Module impacts**

**Extended Data Table 2 | Fractured model setting**

**Extended Data Fig. 1 | Model geometry and computational mesh for Case 3. a,** Fractures are 1 km long and distributed within the upper 12 km. Two-dimensional domain with a surface arc length of 1,000 km and a thickness of 80 km. The yellow star, blue triangles and red marks indicate the source, receivers and fractures, respectively. The dashed blue line marks the inner PML boundary. Boxes indicate the regions enlarged in b–d. **b,** Near-surface mesh, with local refinement around fractures represented as vacuum cavities (white). **c,d,** Mesh near the lateral boundary (c) and lower corner (d), showing the 6 km thick perfectly matched layers (PMLs; red mesh). **e,** Fracture counts in successive depth intervals (cyan bars), with the blue curve indicating the decreasing trend with depth.

**Extended Data Fig. 2 | Comparison of simulated and Apollo seismic envelopes for Cases 2 and 3 at 0.30–0.80 Hz.** Normalized seismic envelopes from Apollo observations (black) are compared with the corresponding synthetic envelopes for Case 2 (red) and Case 3 (blue) for 13 impact–station pairs spanning epicentral distances from 2.2° to 28.0°. **a,** Comparisons for the three shortest-distance records (2.2°, 2.4°, and 3.0°). **b,** Comparisons for the remaining ten records (3.8°–28.0°). For each impact–station pair, the upper and lower panels show Case 2 and Case 3, respectively, with the corresponding Apollo envelope overlaid in black. r, Pearson envelope correlation over 0–4500 s.

**Extended Data Fig. 3 | Determination of envelope peak time and coda decay time for the Apollo observation and numerical simulations. a–c,** Normalized seismic envelopes for the 6.1° Apollo 15 S-IVB–S14 record (a), Case 2 simulation (b), and Case 3 simulation (c), respectively. All records were band-pass filtered at 0.40–0.80 Hz. The gray curves show envelopes smoothed using a moving average with a 25-s half-window, whereas the black curves show the strongly smoothed envelopes used to determine the envelope peak time, $T_{max}$, and the coda decay time, $\tau_d$. The strong smoothing uses a moving-average half-window corresponding to 150 periods at the central frequency (250 s; total window ≈ 500 s). The green dashed line marks $T_{max}$, defined as the time of the maximum strongly smoothed envelope within 0–4500 s. The blue shaded interval denotes the decay-fitting window, which begins 200 s after $T_{max}$ and extends for 1000 s. Within this interval, the

exponential fit curve $E = Ae^{-\frac{t}{\tau_d}}$ obtained by fitting the envelope decay is indicated in red. Envelopes and fitted curves are normalized by the corresponding peak envelope amplitude for visualization. The displayed time window is 0–3000 s.

**Extended Data Fig. 4 | Effects of PDART preprocessing and manual treatment of anomalous intervals on Apollo seismic waveforms and envelopes. a,** Comparison of the original (top, black) and PDART-processed record (bottom, blue) for the Apollo 17 S-IVB impact recorded at station S16 (epicentral distance, 28.0°). **b,c,** Comparison between the PDART-processed records (top, blue) and the corresponding records after manual treatment (bottom, orange) of anomalous intervals for the Apollo 12 LM impact recorded at S12 (2.4°) and the Apollo 17 LM impact recorded at S15 (25.4°), respectively.

**Extended Data Fig. 5 | Sensitivity of Apollo–synthetic envelope correlations to fracture-width variations.** Pearson envelope correlation coefficients between Apollo observations and synthetic seismograms are shown as a function of epicentral distance for the reference Case 2 model (black) and two models in which the fracture-width parameter is decreased (Narrow 50%; blue) or increased (Wide 50%; red).

**Extended Data Fig. 6 | Sensitivity of observed–synthetic envelope agreement to the Gaussian smoothing width. a–c,** Comparison of the Apollo envelope (black) and Case 2 synthetic envelope (red) for the Apollo 17 S-IVB impact recorded at station S12 (epicentral distance, 11.1°), using Gaussian smoothing full widths at half maximum (FWHMs) of 25 s (a), 20 s (b) and 30 s (c). r, Pearson envelope correlation over 0–4500 s. **d,** Mean Pearson r as a function of Gaussian smoothing FWHM across the 13 Apollo impact records. Points denote the arithmetic mean and error bars indicate ±1 sample standard deviation across records. The vertical dashed line at 25 s identifies the smoothing width adopted in the primary analysis.

## Extended Data Figures and Tables

**Extended Data Table 1 | Apollo signals used from Saturn IVB booster and Lunar Module impacts**

| Impact source | Station | Epicentral distance | Component |
|---|---|---|---|
| A14/LM | S14 | 2.2° (65.7 km) | MHZ |
| A12/LM | S12 | 2.4° (73.5 km) | MHZ |
| A15/LM | S15 | 3.0° (92.2 km) | MHZ |
| A14/LM | S12 | 3.8° (115.3 km) | MHZ |
| A13/S-IVB | S12 | 4.5° (135.9 km) | MHZ |
| A16/S-IVB | S12 | 5.1° (153.9 km) | MHZ |
| A14/S-IVB | S12 | 5.8° (175.4 km) | MHZ |
| A15/S-IVB | S14 | 6.1° (185.5 km) | MHZ |
| A16/S-IVB | S14 | 9.1° (274.5 km) | MH1 |
| A17/S-IVB | S12 | 11.1° (337.5 km) | MHZ |
| A15/S-IVB | S12 | 11.7° (355.3 km) | MHZ |
| A17/LM | S15 | 25.4° (771.1 km) | MHZ |
| A17/S-IVB | S16 | 28.0° (850.3 km) | MHZ |

## Extended Data Table 2 | Fractured model setting

| Case | Model configuration | Fracture-bearing depth range, $z$ (m) | Fracture-density parameter, $\varepsilon$ | Fracture length, $L$ (m) |
|---|---|---|---|---|
| 1 | Reference model without fractures. | 0-12000 | 0 | 1000 |
| 2 | Fractures confined to the uppermost 1 km. | 0-1000 | 0.025 | 1000 |
| 3 | Fractures distributed throughout the uppermost 12 km. | 0-12000 | 0.025 | 1000 |
| 4 | Fracture-free uppermost 1 km; fractures restricted to depths of 1–2 km. | 1000-2000 | 0.025 | 1000 |
| 5 | Fracture-free uppermost 1 km; fractures distributed from 1 to 12 km depth. | 1000-12000 | 0.025 | 1000 |
| 6 | Fracture-free uppermost 3 km; fractures restricted to depths of 3–8 km. | 3000-8000 | 0.025 | 1000 |
| 7 | Fracture-free uppermost 2 km; fractures distributed from 2 to 12 km depth. | 2000-12000 | 0.025 | 1000 |
| 8 | Same depth interval and fracture-density parameter as Case 2, with fracture length reduced by 50%. | 0-1000 | 0.025 | 500 |
| 9 | Fracture length doubled relative to Case 2. The fracture-bearing interval was extended to the uppermost 2.5 km to accommodate 2-km-long fractures. | 0-2500 | 0.025 | 2000 |
| 10 | Fracture length increased to 4 km. The fracture-bearing interval was extended to the uppermost 4.5 km to accommodate the longer fractures. | 0-4500 | 0.025 | 4000 |
| 11 | Same geometry as Case 2, with the fracture-density parameter reduced by 50%. | 0-1000 | 0.0125 | 1000 |
| 12 | Same geometry as Case 2, with the fracture-density parameter doubled. | 0-1000 | 0.05 | 1000 |

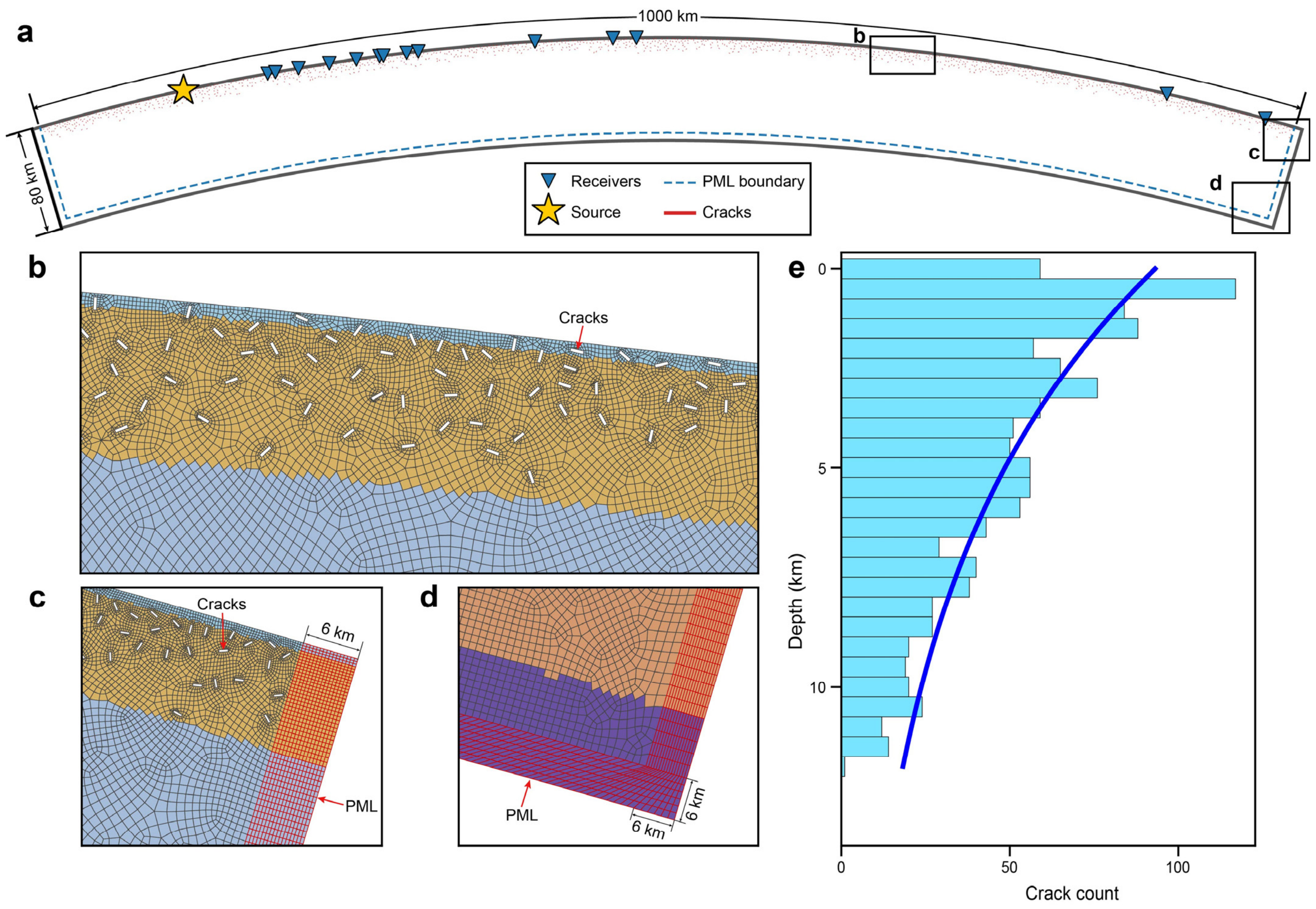

**Extended Data Fig. 1 | Model geometry and computational mesh for Case 3. a,** Fractures are 1 km long and distributed within the upper 12 km. Two-dimensional domain with a surface arc length of 1,000 km and a thickness of 80 km. The yellow star, blue triangles and red marks indicate the source, receivers and fractures, respectively. The dashed blue line marks the inner PML boundary. Boxes indicate the regions enlarged in b–d. **b,** Near-surface mesh, with local refinement around fractures represented as vacuum cavities (white). **c,d,** Mesh near the lateral boundary (c) and lower corner (d), showing the 6 km thick perfectly matched layers (PMLs; red mesh). **e,** Fracture counts in successive depth intervals (cyan bars), with the blue curve indicating the decreasing trend with depth.

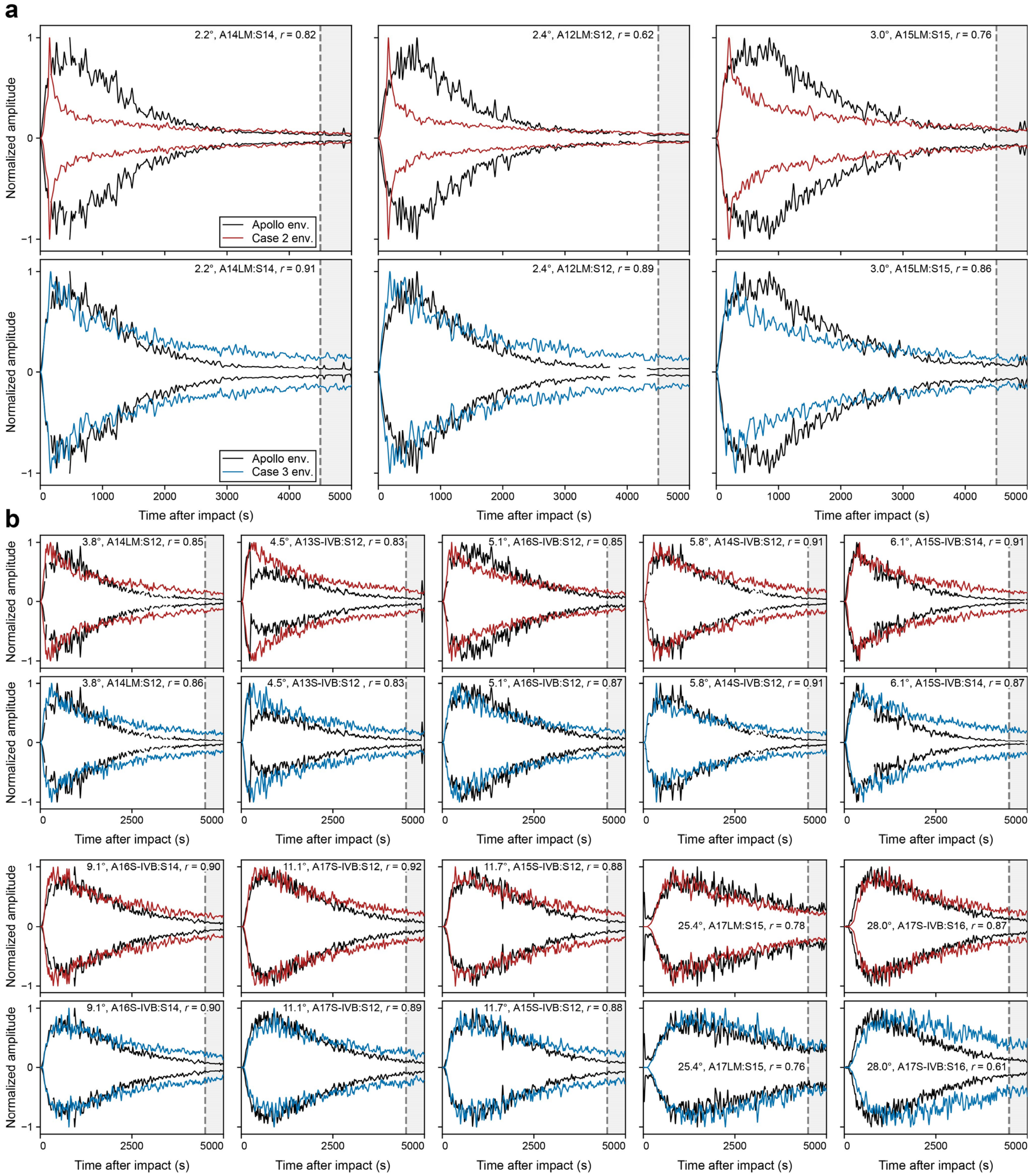

**Extended Data Fig. 2 | Comparison of simulated and Apollo seismic envelopes for Cases 2 and 3 at 0.30–0.80 Hz.** Normalized seismic envelopes from Apollo observations (black) are compared with the corresponding synthetic envelopes for Case 2 (red) and Case 3 (blue) for 13 impact–station pairs spanning epicentral distances from 2.2° to 28.0°. **a,** Comparisons for the three shortest-distance records (2.2°, 2.4°, and 3.0°). **b,** Comparisons for the remaining ten records (3.8°–28.0°). For each impact–station pair, the upper and lower panels show Case 2 and Case 3, respectively, with the corresponding Apollo envelope

overlaid in black. r, Pearson envelope correlation over 0–4500 s.

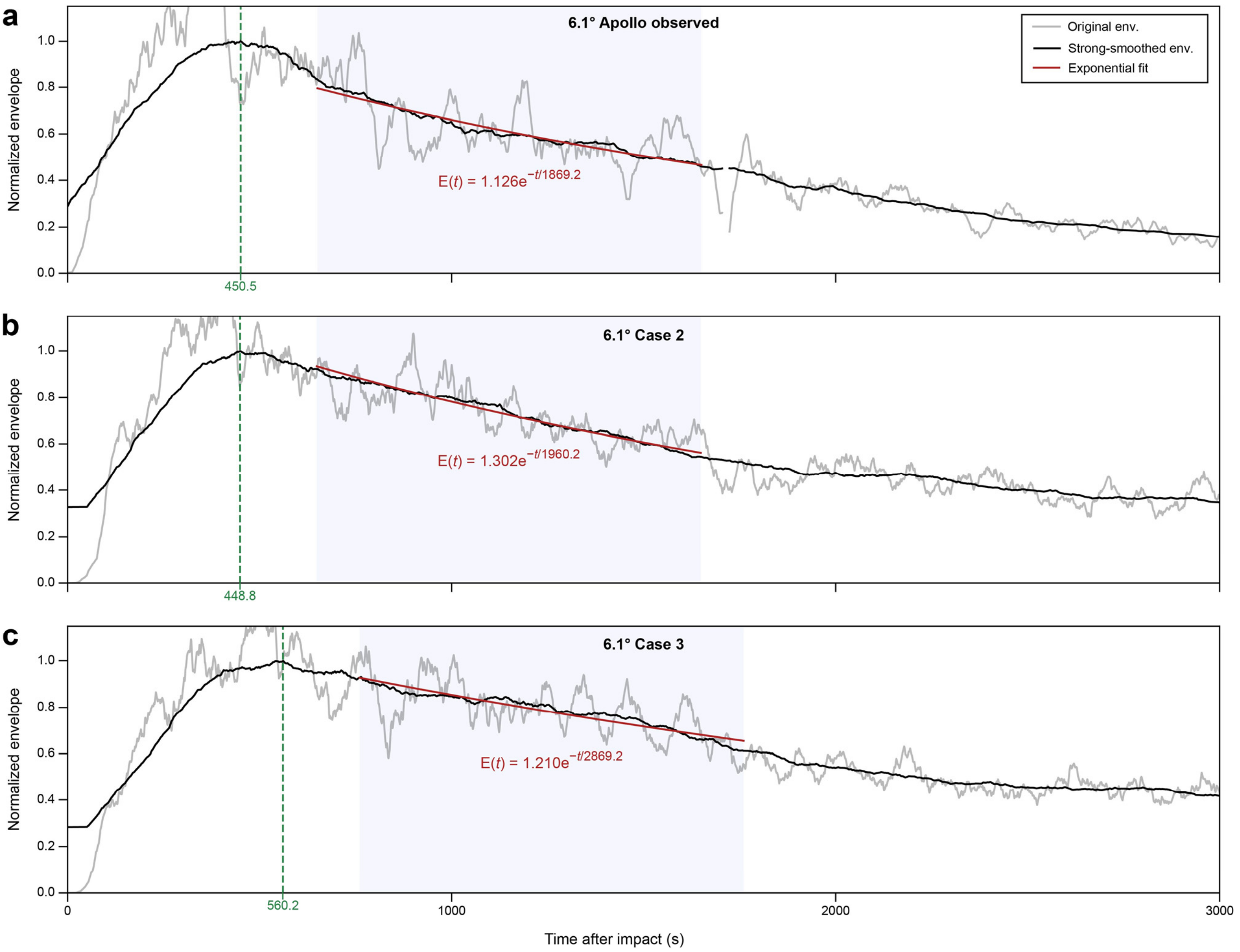

**Extended Data Fig. 3 | Determination of envelope peak time and coda decay time for the Apollo observation and numerical simulations. a–c**, Normalized seismic envelopes for the 6.1° Apollo 15 S-IVB–S14 record (a), Case 2 simulation (b), and Case 3 simulation (c), respectively. All records were band-pass filtered at 0.40–0.80 Hz. The gray curves show envelopes smoothed using a moving average with a 25-s half-window, whereas the black curves show the strongly smoothed envelopes used to determine the envelope peak time, $T_{max}$, and the coda decay time, $\tau_d$. The strong smoothing uses a moving-average half-window corresponding to 150 periods at the central frequency (250 s; total window ≈ 500 s). The green dashed line marks $T_{max}$, defined as the time of the maximum strongly smoothed envelope within 0–4500 s. The blue shaded interval denotes the decay-fitting window, which begins 200 s after $T_{max}$ and extends for 1000 s. Within this interval, the exponential fit curve $E = Ae^{-\frac{t}{\tau_d}}$ obtained by fitting the envelope decay is indicated in red. Envelopes and fitted curves are normalized by the corresponding peak envelope amplitude for visualization. The displayed time window is 0–3000 s.

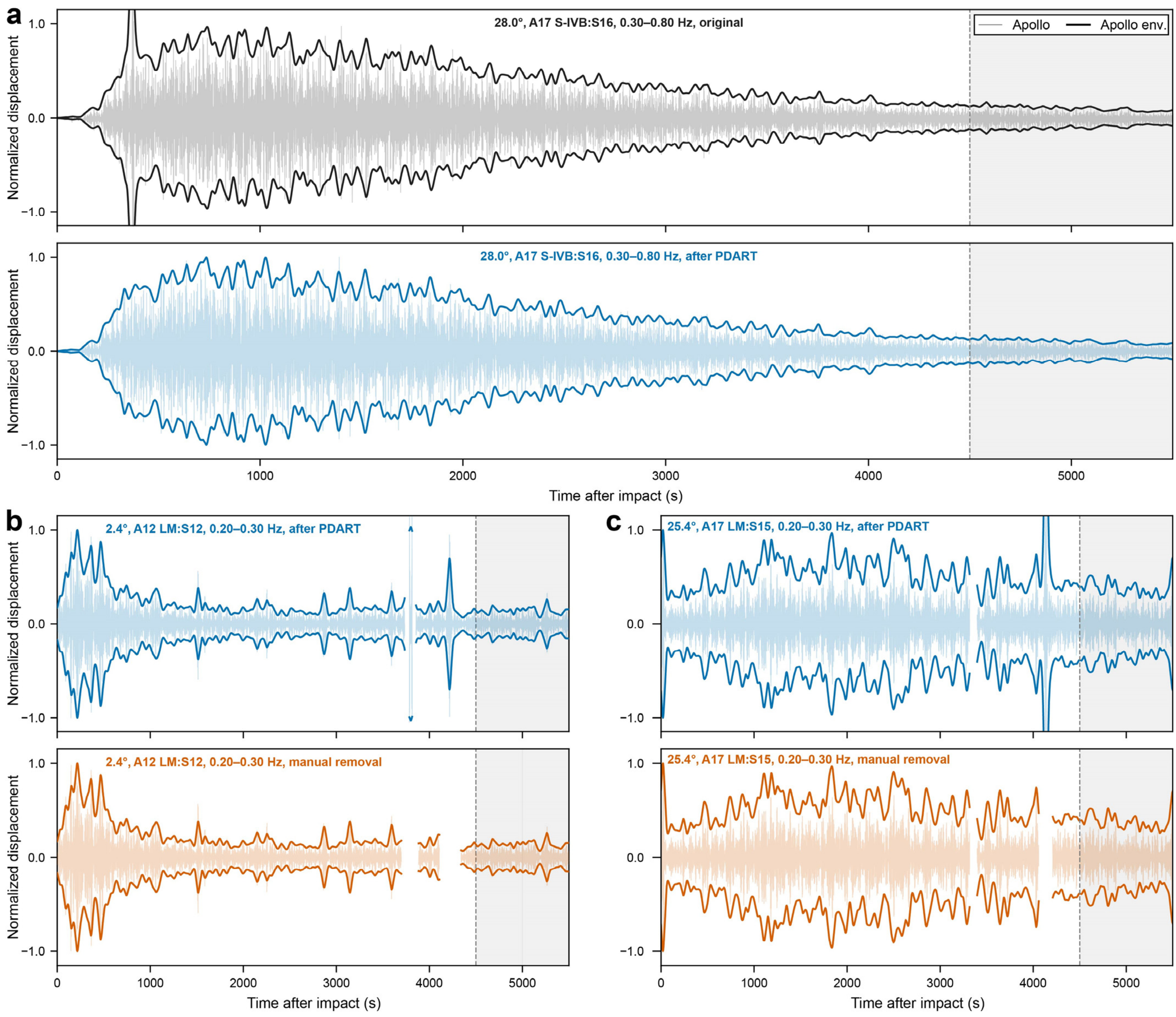

**Extended Data Fig. 4 | Effects of PDART preprocessing and manual treatment of anomalous intervals on Apollo seismic waveforms and envelopes. a**, Comparison of the original (top, black) and PDART-processed record (bottom, blue) for the Apollo 17 S-IVB impact recorded at station S16 (epicentral distance, 28.0°). **b,c,** Comparison between the PDART-processed records (top, blue) and the corresponding records after manual treatment (bottom, orange) of anomalous intervals for the Apollo 12 LM impact recorded at S12 (2.4°) and the Apollo 17 LM impact recorded at S15 (25.4°), respectively.

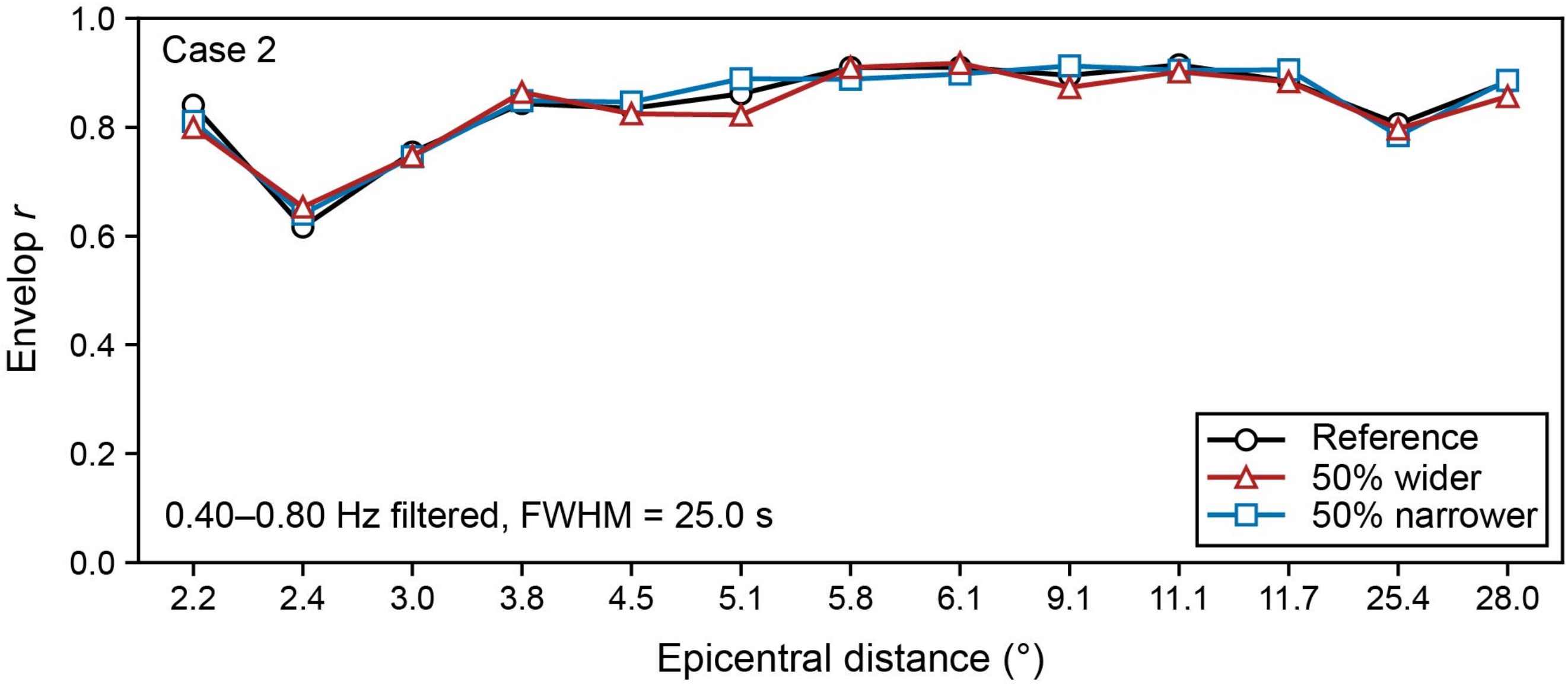

**Extended Data Fig. 5 | Sensitivity of Apollo–synthetic envelope correlations to fracture-width variations.** Pearson envelope correlation coefficients between Apollo observations and synthetic seismograms are shown as a function of epicentral distance for the reference Case 2 model (black) and two models in which the fracture-width parameter is decreased (Narrow 50%; blue) or increased (Wide 50%; red).

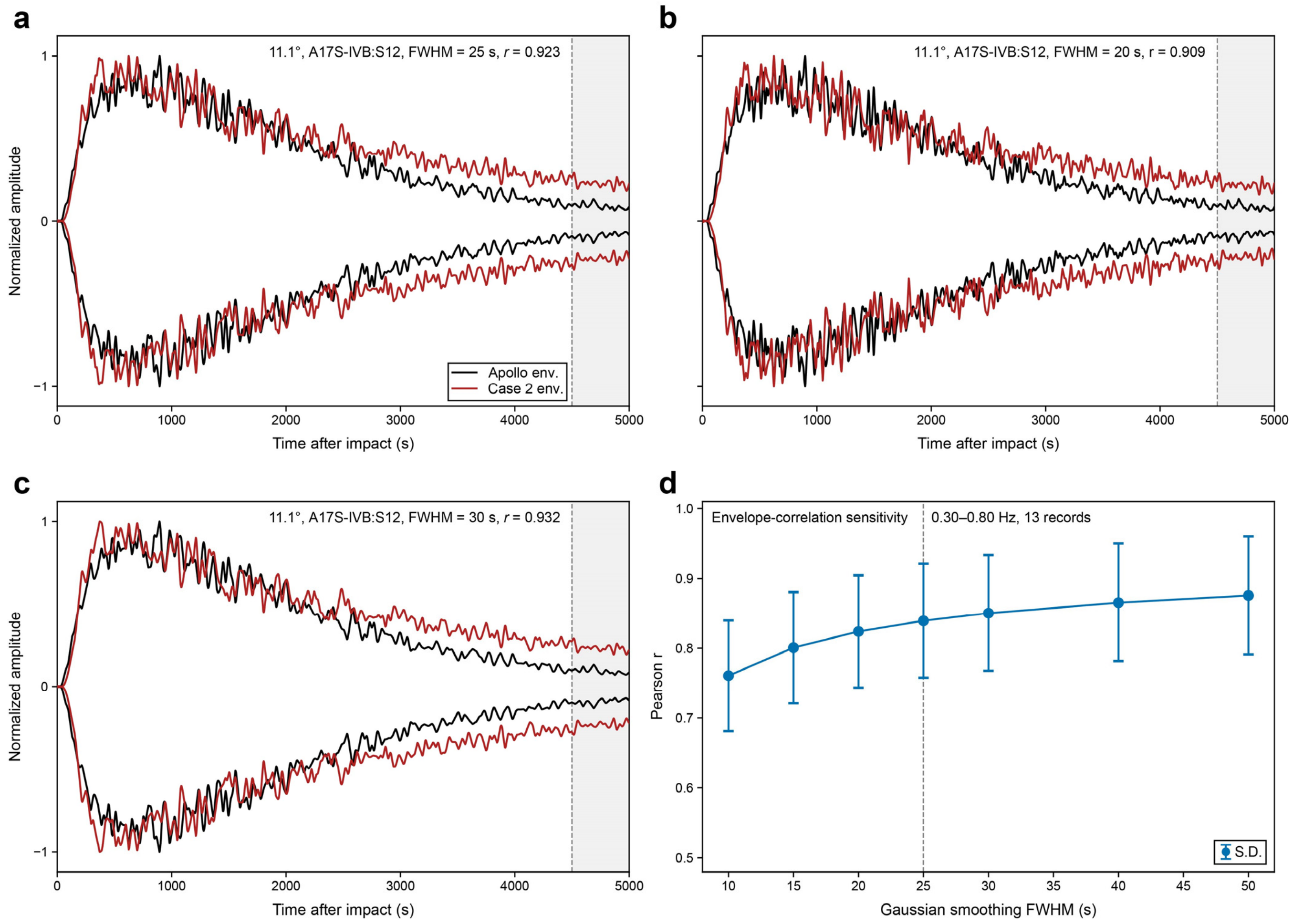

**Extended Data Fig. 6 | Sensitivity of observed–synthetic envelope agreement to the Gaussian smoothing width. a–c,** Comparison of the Apollo envelope (black) and Case 2 synthetic envelope (red) for the Apollo 17 S-IVB impact recorded at station S12 (epicentral distance, 11.1°), using Gaussian smoothing full widths at half maximum (FWHMs) of 25 s (a), 20 s (b) and 30 s (c). r, Pearson envelope correlation over 0–4500 s. **d,** Mean Pearson r as a function of Gaussian smoothing FWHM across the 13 Apollo impact records. Points denote the arithmetic mean and error bars indicate ±1 sample standard deviation across records. The vertical dashed line at 25 s identifies the smoothing width adopted in the primary analysis.